\documentclass[aps,prd,twocolumn,showpacs,groupedaddress]{revtex4-1}

\begin{document}

\title{Finding patterns within perturbative approximation in QCD and indirect relations}

\author{M. R. Khellat}
\email{m.khellat@gmail.com}
\affiliation{Physics Department, Yazd University, 89195-741, Yazd, Iran\\}

\date{\today}

\begin{abstract}
We make an extensive use of BLM approach to study details of predicting higher order corrections based on the approach. This way we are able to test two 
procedures to improve the prediction process. Beside the main line of the two procedures it is found out that overall normalization could change BLM 
patterns effectively. Finally we try to find out whether a BLM pattern is sufficient for a prediction or not, and how one should use such a pattern.
\end{abstract}

\pacs{11.15.Bt 12.38.Bx 11.90.+t}

\maketitle


\section{\label{sec-intro}Introduction}
  Optimization procedures provide a good framework in investigating the behavior of perturbative descriptions of gauge field theories' observables. 
Perturbative analysis of these theories encounters the problem of scheme-scale ambiguity. An appropriate approach to this problem should give warnings
 in cases where it should not be applied, or we should have a mechanism to check the validity. BLM approach \cite{Brodsky:1982gc} contains such a
 mechanism, it is not applicable in analyzing processes containing gluon-gluon interactions in leading order since we should not absorb all the 
flavor dependent part into the running coupling constant in these cases. If we use the approach to resolve scheme ambiguity and find 
physical schemes, such processes will automatically be eliminated. Strange perturbative series would be the outcome of using them as the physical scheme.
 The mathematical form of perturbative series suggests a kind of symmetry and relation between observables but without considering 
optimization processes the real hidden relation can not be revealed. BLM approach is particularly able to reveal commensurate scale relations
 \cite{Brodsky:1994eh} between observables. This flexibility of BLM approach can provide us with very simple relations such as the 
generalized Crewther relation \cite{Brodsky:1995tb} which can be used to test some serious aspects of QFT such as the violation of 
conformal symmetry due to the renormalization procedure.

One can use optimization procedures to predict higher-order coefficients. Clearly any kind of prediction strongly assumes existence of patterns. 
A pattern at least must not have mathematical defects. However the quality of a proposed pattern is usually determined through the comparison of the
 predicted and exact coefficients. For a flexible approach this comparison could be a guideline in improving the approach. As a matter of fact what 
should be followed here rather than improving an approach is finding patterns. Fortunately we do not have many choices for the first step. The first step
 towards these aims is to make the prediction rebuild exact values. This way we encounter a new face of the observable. This mathematical adaptation
 can be interpreted as isolating parts of the observable. These isolated parts carry the physical considerations which supports the 
optimization procedure as parts of them. The set of isolated parts can be called a pattern. However this is not the only way to find patterns. Regarding 
the approach extension and improvement, we introduce biased BLM approaches. Apparently a normal BLM approach cannot correctly analyze 
gluon-gluon interactions in leading order. Here the suggestion of finding BLM generated parts(the isolated parts generated through BLM 
prediction comparison) inside the observable and using the new obtained parameters instead of the original ones constitutes an immediate biasing.
 Also we introduce a second-class biasing where the leading-order is totally dismissed from prediction. 

From a more illuminative point of view, we check the potential of the approach in revealing a pattern for the observable. In the case of BLM this 
has been preferred to finding a way to separate gluon-gluon interactions from the vacuum polarizations. If the approach can be adjusted to 
find a satisfying pattern, it has jumped over the obstacle.

Following the route of finding patterns will lead us through a pile of equations to a simple sufficient condition for the relations that 
connects observables. The condition is a sufficient one if someone wants to get new results from prediction through a second observable. 

For the convenience of the reader: firstly in this way of finding patterns and biasing we deal with systems of equations that are built upon
 comparisons. These comparisons may aim different types of parameters in the pattern such as the perturbative series coefficients or solely the local 
symmetry parameters. Secondly we make notices anywhere one may find himself in a state to perform a selection of equations or parameters 
such as when dealing with unequal number of equations and parameters.  

In the beginning of Sec.\ref{sec-procedure} we will discuss primary aspects of prediction. Deviation pattern approach is introduced in 
Sec.\ref{sec-procedure1} accompanied by its details and uses for Higgs boson decay widths, Bjorken sum rule and Adler function and also 
some complementary discussions on prediction. In this section we also consider patterns developed from the combination of different patterns.
 As we go further we will try take advantage of a second observable for finding patterns in Sec.\ref{sec-2nd proced} and introduce indirect relations.

\section{\label{sec-procedure}The procedures}
We consider a RG-invariant perturbative expansion $R(Q^2)=\;r_0+r_1 a_s(Q^2)+\sum_{i=0}^\infty r_i {a_s}^i(Q^2)$ where $a_s=\frac{\alpha_s}{\pi}$
 ($\alpha_s$ is the RG-invariant effective coupling constant) and $r_i=\sum_{k=0}^{i-1} r_{ik} n_f^k$ for any effective charge or MS-like scheme.
 Predicting any coefficient $r_i$ using BLM approach results in a series of number of flavors $n_f$ for that order that depends on all the lower order
 coefficients. For instance the predicted third order coefficient is
   \begin{eqnarray}\label{r3pre}
   r_{3}^{pre}=&&-\frac{121}{16}\frac{C_{A}^2}{T_F^2}\frac {r_{21}^2}{r_1}\;-\;\frac{17}{8}\frac{C_{A}^2}{T_F}r_{21}
   \;-\; \frac{11}{2}\frac{C_{A}}{T_F}\frac{r_{20} r_{21}}{r_1}\nonumber\\
   &&+\;(2 \frac{r_{20} r_{21}}{r_1} + \frac{5}{4} r_{21} C_{A} + \frac{3}{4} r_{21} C_{F}) n_f + \frac{r_{21}^2} {r_{1}} n_f^2.\;\;\;\;\;
   \end{eqnarray}
Here we have used the 4-dimensional $\beta$-function of the massless MS-like schemes. The $n$th order predicted coefficient 
 $r_{n}^{pre}=\sum_{k=0}^{n-1} r_{nk}^{pre} n_f^k$ appears when re-expanding
 the BLM scale-fixed $(n-1)$th order series
   \begin{eqnarray*}\label{BLM sfs}
  R = r_0 + r_1 a({Q^*}^2) + \bar{r}_2 a^2({Q^*}^2) + \ldots + \bar{r}_{n-1} a^{n-1}({Q^*}^2).
   \end{eqnarray*}
$Q^*$ is to absorb all vacuum polarization. Re-expanding the above series happens through the $n$th order running of the couplant:
   \begin{eqnarray*}\label{rc-nth order}
  a({Q^*}^2)=a(Q^2) - a^2(Q^2)\beta_0\ell + a^3(Q^2)(\beta_0^2\ell^2 - \beta_1 \ell)\;\;\;\;\;\;\;\;\;\;\;\;\;\\
   + a^4(Q^2) (-\beta_0^3\ell^3 + \frac{5}{2} \beta_0\beta_1\ell^2 - \beta_2 \ell) + \ldots +{\cal O}(a^{n+1}),\;\;\;\;
   \end{eqnarray*}
where $\ell=\ln({Q^*}^2/Q^2)$ and it may occur in multiple steps with appropriate $\ell$s for a multi-scale version of the BLM scale-fixed series.
 We will discuss more details of BLM extensions later.
  
One more preliminary review on overall normalization and Casimir operators is needed before going on to the procedures. Local gauge invariance 
constructs Yang-Mills theories i.e., non-Abelian gauge theories. Such local symmetries suites well to compact semi-simple Lie algebras. For such an 
algebra one can find the overall normalization \cite{vanRitbergen:1997va,Cvitanovic:1976am,Mojaza:2010} that can determine the structure of the algebra 
and its representations. The tensor product of generators of the color $SU(N)$s of two particles,  $T_{a}^A T_{a}^B$, gives the taste of color
 interaction between them. This operator is surprisingly an invariant of the tensor product space. The Casimir operator $T_a^2$, where $T_a$s are
 generators of the product space, related to this interaction is a number depending on the dimension of the representation. $C_A$ and $C_F$ are quadratic
 Casimir operators of the adjoint and fundamental representations transforming gauge and fermion fields. The representation's generators scale factor
 i.e., the trace normalization of the representation is $T_F$ for the fundamental representation. Obviously setting $N_A$, the number of generators of
 the group, $C_F$, $T_F$, and $d_F^{abcd}$, the higher order group invariants, to $1$ and the parameters of the adjoint representation to $0$ will
 separate the $U(1)$ factor of the group. The dynamics of the quenched approximation of the non-Abelian part is revealed in the limit $n_f=0$ for all
 physical energies while the conformal invariant partner of the symmetry group sits where the $\beta$-function is $0$.

\subsection{\label{sec-procedure1}Deviation}
If BLM prediction pattern is some good pattern for the perturbative series, one can rewrite the whole series in terms of just $r_1$ and $r_2$
 coefficients. The quality of the pattern will depend on these parameters beside the BLM approach itself or truthfully the mathematical mechanism 
employed for it. Anyway if BLM approach is able to discover the pattern, it will not be so surprising if the original $r_1$ and $r_2$ do not be the ones
 that develop the pattern. A straightforward step would be to borrow the $r_1$ and $r_2$ that form the real $r_3$ since it is the most important part of
 the pattern. Yet the procedure would be flexible enough to be able to handle exceptional but important cases such as the ones with large higher-order
 corrections. We could discuss the situation also as if the weights of $r_1$ and $r_2$ parts should be other values in Eq.(\ref{r3pre}) or other
 combinations of weights of lower-order coefficients are appropriate.

In general a BLM pattern contains several coupled equations corresponding to all we know about the perturbative series. These equations are acquired
 by comparing the predicted coefficient and the exact ones. For instance when one knows $r_4$ in addition to $r_3$ and wants to adapt the prediction
 pattern so that all the details are involved in it, he will notice that $r_{4}^{pre}$ depends on all the $6$ parts of lower order coefficients
$\{ r_1,r_{20},r_{21},r_{30},r_{31},r_{32} \}$. But the 4th order comparison($r_{4}^{pre}=r_4$) yields $4$ equations. Considering the 3rd order 
comparison, we see that $7$ equations are controlled by just $6$ variables. This indicates that one of these $7$ parts is generated automatically 
through the adaptation of the other parts. The more information, the more automatic generation. So here one is forced to do a selection of equations 
considering the importance of the corresponding comparison. For instance one may find out that $r_{43}^{pre}$ plays the least important role in the 
deviation of $r_{4}^{pre}$ from $r_4$, the comparison $r_{43}^{pre}=r_{43}$ will be easily removed from the list of equations. The same works for other
selections of parameters too. Deviation Pattern procedure obtains new parameters through a selection of comparisons and then replaces the original
 parameters with the new ones to build the higher-order coefficients. 

It is possible to involve Adler transformation into the above procedure by performing the prediction or comparison for the corresponding Adler function. 
 Some notes regrading perturbative aspects of Adler function will follow. The real function $R(s)$ where $s$ denotes the center-of-mass energy squared is 
related to the Adler function $D(Q^2)$ through
\begin{eqnarray*}\label{Adler trans.}
D(Q^2)=Q^2\;\int_0^{\infty} \frac{\mathrm{d}s}{(s + Q^2)^2} R(s).
\end{eqnarray*}
The reverse transformation rebuilds $R(s)$ as follows:
\begin{eqnarray*}\label{rev Adler trans.}
R(s)=\frac{i}{2\pi}\;\int_{s - i\epsilon}^{s + i\epsilon} \frac{\mathrm{d}z}{z} D_{pt}(-z).
\end{eqnarray*} 
One may use the RG-invariant perturbative approximation of the Adler $D$-function $D_{pt}(Q^2)=\sum_n d_n \alpha_s^n(Q^2)$ in the reverse transformation.
This will generate the time-like RG-invariant effective coupling $\tilde{\alpha}(s)$ and a set of pipizated functions. Certainly we deal with the 
analytical continuations \cite{Shirkov:1996cd} of Euclidean perturbative expansions when using the reverse transformation. This results in pipizated 
expressions \cite{Shirkov:2000db,Shirkov:2000qv}. The non-power sets of pipizated functions and analyticized functions replace time-like and space-like 
couplings which regenerate the perturbative series in terms of a self-consistent scheme.

\subsubsection{\label{sec-Hdecay}Higgs decay widths}
As an important and interesting illustration of the Deviation Pattern procedure (DPA), we first consider Higgs boson decays \cite{Djouadi:2005gi}
 into bottom quarks \cite{Baikov:2005rw,Kataev:2008ym} and gluons \cite{Baikov:2006ch}. The hadronic decay width of the Higgs boson seems to have large
 QCD corrections to the Born approximation. Considering $M_H>>2m_q$ results in
\begin{eqnarray*}\label{H2.light quarks}
\Gamma(H \rightarrow q\bar{q})= \frac{3G_F}{4 \sqrt{2} \pi M_H} m_q^2 \tilde{R}(s)
\end{eqnarray*} 
for the decay width, where the term to the left of $\tilde{R}$ is $\Gamma_{Born}(H \rightarrow q\bar{q})$. The absorptive part of the corresponding two
 point correlator $\tilde{R}$ generates the QCD corrections. Gluonic decay width takes the following form based on the fact that top quarks plays the
 main role in coupling of gluons to Higgs boson,
\begin{eqnarray*}\label{H2.gg width}
\Gamma(H \rightarrow gg)= \frac{\sqrt{2} G_F}{M_H} C_1^2 \mathrm{Im} \Pi^{GG}(q^2),
\end{eqnarray*}
where $C_1$ carries the top mass dependence and $\Pi^{GG}$ is the induced vacuum polarization by the renormalized gluon operator. The gluonic decay width
 is factorized by the $K$-factor, $\Gamma(H \rightarrow gg)=K \Gamma_{Born}(H \rightarrow gg)$.

Now we are in a position that we should consider generalizations of the BLM approach. For a historical overview of the extensions to the approach see 
\cite{Mikhailov:2004iq} and references therein. Our emphasis here is on the single-scale extension \cite{Grunberg:1991ac} used in
 \cite{Brodsky:1995tb} and the multi-scale one developed in \cite{Brodsky:1994eh}. Scale or scales are in need of $n(n-1)/2$ parameters to be able to
 absorb all vacuum polarization insertions in the $n$th order truncated series. The single-scale extension uses these parameters at once. However the
 multi-scale version takes them step by step with a specific theoretical consideration. Categorizing corrections forms the basis of some of the other
 generalizations such as \cite{Mikhailov:2004iq} also. Taking advantage of the single-scale $\ell$ results in 
\begin{widetext}
\begin{eqnarray}\label{r4pre.ss}
r_{4}^{pre} =&& - \frac{187}{32} \frac{C_A^3}{T_F^2} r_{32} - \frac{17}{8} \frac{C_A^2}{T_F} r_{31} + \frac{203}{1152} \frac{C_A^3}{T_F} r_{21} 
+ \frac{51}{32} \frac{C_A^2 C_F}{T_F} r_{21} - \frac{121}{8} \frac{C_A^2}{T_F^2} \frac{r_{20}}{r_1} r_{32}
- \frac{3993}{64} \frac{C_A^3}{T_F^3} \frac{r_{21}}{r_1} r_{32} - \frac{11}{2} \frac{C_A}{T_F} \frac{r_{20}}{r_1} r_{31}\nonumber\\
&&  - \frac{363}{16} \frac{C_A^2}{T_F^2} \frac{r_{21}}{r_1}r_{31} + 11 \frac{C_A}{T_F} \frac{r_{20}^2}{r_1^2} r_{21} 
+ \frac{55}{8}  \frac{C_A^2}{T_F} \frac{r_{20}}{r_1}r_{21} + \frac{33}{8} \frac{C_A C_F}{T_F} \frac{r_{20}}{r_1} r_{21} 
- \frac{33}{4} \frac{C_A}{T_F} \frac{r_{30}}{r_1} r_{21} + \frac{605}{16} \frac{C_A^2}{T_F^2} \frac{r_{21}^2}{r_1^2} r_{20}\nonumber\\
&& + \frac{1331}{32} \frac{C_A^3}{T_F^3} \frac{r_{21}^3}{r_1^2} + \frac{1089}{64} \frac{C_A^2 C_F}{T_F^2} \frac{r_{21}^2}{r_1} 
+ \frac{627}{32} \frac{C_A^3}{T_F^2} \frac{r_{21}^2}{r_1} 
+ [ (\frac{5}{4} C_A + \frac{3}{4} C_F) r_{31} + \frac{515}{576} C_A^2 r_{21} - \frac{155}{192} C_A C_F r_{21}  - \frac{21}{32} C_F^2 r_{21}\nonumber\\
&&  + \frac{21}{16} \frac{C_A^2}{T_F} r_{32} + \frac{33}{16} \frac{C_A C_F}{T_F} r_{32} + \frac{121}{16} \frac{C_A^2}{T_F^2} \frac{r_{21}}{r_1} r_{32} 
+ \frac{11}{4} \frac{C_A}{T_F} \frac{r_{21}}{r_1} r_{31} - \frac{11}{2} \frac{C_A}{T_F} \frac{r_{21}^2}{r_1^2} r_{20}
- \frac{173}{32} \frac{C_A^2}{T_F} \frac{r_{21}^2}{r_1} 
- \frac{5}{2} C_A \frac{r_{20} r_{21}}{r_1}\nonumber\\
&&- \frac{3}{2} C_F \frac{r_{20} r_{21}}{r_1} - \frac{121}{16} \frac{C_A^2}{T_F^2} \frac{r_{21}^3}{r_1^2} 
- \frac{165}{32} \frac{C_A C_F}{T_F} \frac{r_{21}^2}{r_1}+ 2 \frac{r_{20} r_{31}}{r_1} - 4 \frac{r_{20}^2 r_{21}}{r_1^2} 
+ 3 \frac{r_{30} r_{21}}{r_1} ] n_f \nonumber\\
&& + [ (\frac{5}{4} C_A + \frac{3}{4} C_F) r_{32} - (\frac{79}{288} C_A T_F + \frac{11}{48} C_F T_F) r_{21}  
- (\frac{5}{8} C_A - \frac{3}{8} C_F) \frac{r_{21}^2}{r_1} + 2 \frac{r_{21} r_{31}}{r_1} - 3 \frac{r_{21}^2 r_{20}}{r_1^2}
+ 2 \frac{r_{20} r_{32}}{r_1} ] n_f^2 \nonumber\\
&& + (-\frac{r_{21}^3}{r_1^2} + 2 \frac{r_{21} r_{32}}{r_1}) n_f^3,
\end{eqnarray}
\end{widetext}  
as the predicted 4th order coefficient. Strange thing is the appearance of $T_F$ in the denominator of Eq.(\ref{r3pre}) and Eq.(\ref{r4pre.ss}). This is
 even more unusual for the multi-scale based prediction where the common factor $r_1^2 (4 T_F r_{20} + 11 C_A r_{21}) T_F^3$ sits as the denominator of
 the whole $r_{4}^{pre}$. Apparently in both cases $r_{4}^{pre}$ does not posses a normal coefficient structure. A normal structure is constructed 
through a combination of $\beta$-coefficients multiplications. For re-expansion we are using a version of $\ell$ which contains $n(n-1)/2$ parameters for 
the $n$th order series. One suggestion might be to add $n$ extra parameters to $\ell$ or $\ell$s when re-expanding to the $(n+1)$th order series and see
 whether a standard structure of $r_{4}^{pre}$ could determine these parameters. Perturbative series parameters $\{ r_1, r_2, r_3 \}$ presented in
 Eq.(\ref{r3pre}) and Eq.(\ref{r4pre.ss}) intrinsically contains local symmetry group parameters $\{ C_A, C_F, T_F, \ldots \}$. So before following the
 above suggestion, relations must be rewritten in terms of new series parameters which are group parameters independent. However we will get around
 this and take advantage of the single-scale extension in the following predictions.

In this way the normal BLM prediction generates the following series as the 4th order coefficient of $\tilde{R}$ comparing to the calculated one:
\begin{eqnarray*}\label{r4pre.Hbb}
\tilde{r}_{4}^{pre} &&= 1103.4 - 268.480 n_f + 10.67 n_f^2 - 0.046 n_f^3, \\
\tilde{r}_{4}\;\;\; &&= 39.354 - 220.943 n_f + 9.685 n_f^2 - 0.020 n_f^3.
\end{eqnarray*}
If we perform comparison and prediction for the corresponding Adler function, we will have:
\begin{eqnarray*}\label{r4pre.Hbb.Adler}
\tilde{r}_{4}^{pre} = -1627.6 + 191.328 n_f - 11.41 n_f^2 + 0.245 n_f^3,
\end{eqnarray*}
which indicates an improvement.

The normal prediction for the 4th order coefficient of the $K$ factor reads:
\begin{eqnarray*}\label{g3pre.Hgg}
g_{3}^{pre} &&= 2219.878 - 901.698 n_f + 53.524 n_f^2- 0.516 n_f^3, \\
g_3 \;\;\; &&= 3372.073 - 866.588 n_f + 48.088 n_f^2 - 0.538 n_f^3.
\end{eqnarray*}
Performing comparison and prediction for the corresponding Adler function improves the result:
\begin{eqnarray*}\label{g3pre.Hgg.Adler}
g_{3}^{pre}= 3783.694 - 792.353 n_f + 42.653 n_f^2 - 0.581 n_f^3,
\end{eqnarray*}
whereas preventing the first term of the perturbative series affect the prediction during re-expansion (a biased BLM prediction \cite{Mirjalili:2010})
 alongside the above procedure results in a better behavior for higher number of flavors:
\begin{eqnarray*}\label{g3pre.Hgg.Biased}
g_{3}^{pre}= 533.831 - 189.814 n_f + 15.738 n_f^2 - 0.318 n_f^3.
\end{eqnarray*}
Biased prediction refers to this kind of biasing. 

Briefly normal BLM prediction for both the decays results in patterns whose deviations from the calculated coefficients do not have serious $n_f$
 dependences. Specially for $\tilde{R}$ it is so clear that the major part of deviation is due to the first part of 4th order coefficient. In the case of
 $K$ factor we have two large parts. The deviation is shared among both of them, however again the first part plays the main role. DPA will make
 $\tilde{r}_{4}^{pre}$ recover $\tilde{r}_{4}$ for higher values of $n_f$, whilst this happens for $g_{3}^{pre}$ when DPA is utilized by a second-class 
biasing. 

If we consider the ratio $a_s(Q^2) \frac{K}{\tilde{R}}$, a normal BLM prediction will indicate the same behavior as before. This is whilst a biased one
 shows a tendency between the predicted and calculated coefficients for higher values of $n_f$. It is more evident in the $\beta$-representation of the 
coefficients:
\begin{eqnarray*}\label{r4pre.Hgg/Hbb}
r_{4pre}^{bia} &&= 81.04 - 259.09 {\beta}_0 + 25.09 {\beta}_0^2 - 15.92 {\beta}_0^3 + 4.03 {\beta}_2, \\
r_4 \;\;\;\; &&= 120.0 - 261.26 \beta_0 + 75.95 \beta_0^2 + 12.96 \beta_0^3 + 4.03\beta_2.
\end{eqnarray*}
The genuine $\beta$-representation and seBLM are developed in \cite{Mikhailov:2004iq} but here we have used a different combination of $\beta$-
coefficients for the $\beta$-representation. As we go further and employ DPA, normal prediction will be closer to the calculated result. So the first
 term in its $\beta$-representation is near $r_4$'s. At the same time the biased prediction decreases its slope and reaches $-54.5961$ for $n_f=6$ in
 comparison to $-382.9029$ of the last step while the calculated value is $-37.2079$. Taking advantage of the Adler function makes the biased prediction 
get away from $r_4$ but the normal prediction continues decreasing its slope and this time gets really close to $r_4$ as $n_f$ increases:
\begin{eqnarray*}\label{r4pre.Hgg/Hbb.Adler}
r_{4}^{pre}= 86.31 - 112.75 \beta_0 + 29.87 \beta_0^2 - 2.49 \beta_0^3 + 4.61 \beta_2,
\end{eqnarray*}
which reads $-35.2114$ for $n_f=6$.

If we establish the ratio $a_s(Q^2) \frac{\tilde{R}}{K}$, normal prediction will generate a result diverging from the calculated coefficient as $n_f$
 increases. As we use DPA it suddenly changes its orientation towards the calculated result. Utilizing DPA with the Adler transformation will make them
 behave more similarly.

\subsubsection{\label{sec-AdlerBjp}Adler function and Bjorken sum rule}
The vacuum polarization induced by vector current $j_{\mu}=\sum_i \bar{\psi_i} \gamma_{\mu} \psi_i$ is the time-ordered correlation function
\begin{eqnarray*}\label{correlation}
(q_{\mu}q_{\nu} - q^2 g_{\mu \nu}) \Pi(Q^2) = i \int {\mathrm{d}}^4 x e^{iq.x} \langle 0 \mid T[j_{\mu}(x) j_{\nu}(0)] \mid 0 \rangle, 
\end{eqnarray*}
where $Q^2=-q^2$. The Adler function corresponding to the current takes form through the scalar correlator $\Pi(Q^2)$,
\begin{eqnarray*}\label{Adler.prod}
D(Q^2)=-12 \pi^2 Q^2 \frac{\mathrm{d}}{\mathrm{d}Q^2} \Pi(Q^2).
\end{eqnarray*}
Integration over polarized proton and neutron structure functions $g_1^{p,n}$ is related to the Bjorken polarized sum rule quantity $d_{Bj}$ through
\begin{eqnarray*}\label{Adler.prod.e}
\Gamma_1^{p-n} &&= \int_{0}^{1} \mathrm{d}x (g_1^p(x,Q^2) - g_1^n(x,Q^2))\\
&&=\frac{g_A}{6} (1 - d_{Bj}(Q^2)) + \sum_{j=2}^{\infty} \frac{\mu_{2j}^{p-n}(Q^2)}{Q^{2j-2}},
\end{eqnarray*}
where $(1-d_{Bj})$ denotes the coefficient function $C_{Bjp}$ and $g_A$ is the charge corresponding to the axial vector current of the nucleon. Recently
 the coefficient function $C^{Bjp}(Q^2)$ and the non-singlet component of the Adler function $D^{NS}(Q^2)$ \cite{Baikov:2010iw} have been calculated to
 order $\alpha_s^4$. They are based on the constraints due to the special form of the generalized Crewther relation $D(a_s)C^{Bjp}(a_s)= 1 +
 \frac{\beta(a_s)}{a_s} \sum_i K_i a_s^i$ \cite{Broadhurst:1993ru,Crewther:1997ux} ($\beta(a_s)$ is the gauge group $\beta$-function) and the color
 structure of the coefficients of these functions \cite{Baikov:2010je}. It is worth notifying that more recently the generalized Crewther relation (CR)
 has been expressed in terms of new $\beta$-independent polynomials by factorizing multiple powers of $\beta$-function \cite{Kataev:2010du}:
\begin{eqnarray}\label{CR.extension}
D(a_s)C^{Bjp}(a_s) = 1 + \sum_{n \geq 1} \bigg( \frac{\beta(a_s)}{a_s} \bigg)^n {\cal P}_n(a_s). 
\end{eqnarray}    
Adapting the old and new structures of CR reveals the structure of ${\cal P}_n$ coefficients.

Prediction results for $C^{Bjp}$ and $D^{NS}$ are summarized in Table~\ref{table:pre1}.%
\begingroup
\squeezetable
\begin{table}[ht]
\caption{\label{table:pre1}$C^{Bjp}$ and $D^{NS}$ prediction results.} 
\begin{ruledtabular}
\begin{tabular}{c c c c c c c c c c c} 
$C^{Bjp}$ & $n_f^0$ & $n_f^1$ & $n_f^2$ & $n_f^3$ & $D^{NS}$ & $n_f^0$ & $n_f^1$ & $n_f^2$ & $n_f^3$ \\ [0.5ex] 
\hline
$c^{pre}_{4}$ & -265.4 & 95.26 & 5.94 & 0.08    & $d^{pre}_{4}$ & 362.1 & -99.08 & 5.04 & -0.05 \\ 
$c^{pre1}_{4}\footnote{DPA.}$ & -261.3 & 67.71 & -3.62 & 0.04  & $d^{pre1}_{4}$ & 317.2 & -93.26 & 4.76 & -0.03  \\
$c^{pre2}_{4}\footnote{DPA utilized by Adler function.}$ & -20.89 & 26.16 & -1.23 & -0.01 & $d^{exact}_{4}$ & 407.4 & -103.3 & 5.63 & -0.03 \\
$c^{exact}_{4}$ & -479.4 & 123.4 & 7.69 & 0.10   \\ [1ex] 
\end{tabular}
\end{ruledtabular}
\end{table}
\endgroup
For $C^{Bjp}$ it seems that a major part of deviation for the normal prediction is due to the first term. But unlike the case of Higgs decay into bottom
 quarks the gap between prediction and calculation is not so large. It is whilst $D^{NS}$ is completely different. The prediction and calculated results
 may seem really close to each other but their $\beta$-representations are really far. There seems to be a similarity between $D^{NS}$'s and Higgs decay
 into gluons width's deviation patterns, both of them have extremum near $n_f=3$. DPA makes an improvement in $C^{Bjp}$ prediction for higher values of
 $n_f$ and provides a slight improvement in the $\beta$-representation. A good improvement in the first term of $\beta$-representation for $D^{NS}$ is
 visible in its DPA prediction. At last taking advantage of Adler transformation will not improve $C^{Bjp}$ prediction.

The results of predictions for the ratio $\frac {D^{NS}(Q^2)} {C^{Bjp}(Q^2)}$ could be illustrated in Table~\ref{table:pre2}.%
\begingroup
\squeezetable
\begin{table}[ht]
\caption{\label{table:pre2}$\frac {D^{NS}} {C^{Bjp}}$ prediction results.} 
\begin{ruledtabular}
\begin{tabular}{c c c c c c} 
$\frac {D^{NS}} {C^{Bjp}}$ & $n_f^0$ & $n_f^1$ & $n_f^2$ & $n_f^3$ & 1st term of the $\beta$-rep. \\ [0.5ex] 
\hline
$dc^{pre}_{4}$   & 1925.9 & -463.57 & 24.11 & -0.2872 & 239.47  \\ 
$dc^{pre1}_{4}$  & 1409.6 & -332.47 & 16.05 & -0.1302 & 397.27  \\
$dc^{pre2}_{4}$  & 111.29 & -140.82 & 7.128 & -0.0043 & 397.27  \\
$dc^{exact}_{4}$ & 2444.4 & -572.69 & 31.02 & -0.3415 & 595.05  \\ [1ex] 
\end{tabular}
\end{ruledtabular}
\end{table}
\endgroup
The deviation pattern of the ratio does not contain any extremum and is decreasing for higher values of $n_f$. DPA prediction is good for $n_f=6$ and
 making use of Adler transformation will not help. It seems that $C^{Bjp}$ in this ratio affects parts of $D^{NS}$ for which the $D^{NS}$
 pattern was weakened while employing DPA. It makes the new predictions better for $n_f=6$.

Table~\ref{table:pre3} shows the results for the ratio $\frac {C^{Bjp}(Q^2)} {D^{NS}(Q^2)}$.
\begingroup
\squeezetable
\begin{table}[ht]
\caption{\label{table:pre3}$\frac {C^{Bjp}} {D^{NS}}$ prediction results.} 
\begin{ruledtabular}
\begin{tabular}{c c c c c c} 
$\frac {C^{Bjp}} {D^{NS}}$ & $n_f^0$ & $n_f^1$ & $n_f^2$ & $n_f^3$ & 1st term of the $\beta$-rep. \\ [0.5ex] 
\hline
$cd^{pre}_{4}$   & -54.66  & 33.49 & -2.26 & 0.0319 & -50.64  \\ 
$cd^{pre1}_{4}$  & -101.41 & 30.14 & -1.67 & 0.0153 & -66.90  \\
$cd^{pre2}_{4}$  & -37.55  & 18.99 & -1.02 & 0.0028 & -66.90  \\
$cd^{exact}_{4}$ & -172.83 & 45.62 & -3.03 & 0.0379 & -150.66  \\ [1ex] 
\end{tabular}
\end{ruledtabular}
\end{table}
\endgroup

Existence of $D^{NS}$ in this ratio absorbs the converging behavior of DPA prediction for higher values of $n_f$. This converts it to a general
 improvement along all values of $n_f$. The corresponding results for the biased BLM prediction are tabulated in Table~\ref{table:pre4}.%
\begingroup
\squeezetable
\begin{table}[ht]
\caption{\label{table:pre4}$\frac {C^{Bjp}} {D^{NS}}$ biased prediction results} 
\begin{ruledtabular}
\begin{tabular}{c c c c c c} 
$\frac {C^{Bjp}} {D^{NS}}$ & $n_f^0$ & $n_f^1$ & $n_f^2$ & $n_f^3$ & 1st term of the $\beta$-rep. \\ [0.5ex] 
\hline
$cd^{pre}_{4}$   & 52.07  &  13.64 & -1.40 & 0.0244 & -72.32   \\ 
$cd^{pre1}_{4}$  & 173.65 & -21.23 &  0.41 & 0      & -140.29  \\
$cd^{pre2}_{4}$  & 461.62 & -73.41 &  3.19 & -0.0409 & -140.29  \\
$cd^{exact}_{4}$ & -172.83 & 45.62 & -3.03 & 0.0379 & -150.66  \\ [1ex] 
\end{tabular}
\end{ruledtabular}
\end{table}
\endgroup
This indicates real improvements in the $\beta$-representation.

\subsubsection{\label{sec-casimir}Quadratic Casimir invariants}
Each part in the first term of a normal BLM pattern Eq.(\ref{r4pre.ss}) contains at least one factor $C_A$. So for the case of Higgs decay into bottom 
quarks setting $C_A=0$ improves the normal prediction significantly along all values of $n_f$ specifically for lower ones.

Working on the ratio $a_s(Q^2) \frac{\tilde{R}}{K}$ and using $\{r_1, r_2\}$ that came out of DPA, one will find out the sensitivity of the $n_f^1$ part
 of the 4th order prediction to $C_A$ in such a way that setting $C_A=0 $, $C_F=1$, and $T_F=1$ results in a $0.04$ relative error in predicting this
 part.

For the ratio $a_s(Q^2) \frac{K}{\tilde{R}}$ an interesting improvement occurs when choosing the canonical i.e., conventional form of the quadratic
 Casimir invariants for $SU(2)$. More freedom is obtained by departing from the conventional choice to the overall normalization
 \cite{vanRitbergen:1998pn} where parameters $b$ and $N$ take control of the normalization. This also indicates that $SU(2)$ is a better choice for this
 ratio.

Adapting the first part of the 4th order coefficient of $C^{Bjp}$ by using the overall normalization indicates that $SU(2)$ improves the $n_f^2$ part and
 $SU(4)$ improves the $n_f^1$ part in comparison to $SU(3)$ while $SU(3)$ still provides a better combination.

A very exciting result is obtained when performing the last procedure for $D^{NS}$. In this case $SU(2)$ gives us such a combination of $n_f^1$ and
 $n_f^2$ parts that makes the BLM pattern fit to the calculated terms in the whole region of $n_f$. It is in spite of the slight separation occurred in 
all the previous cases for higher values of $n_f$.

One should pay attention to these parameters of gauge symmetry because their effects are noticeable on the problem and specially on an approach which
 seems to be related to the concept of conformal symmetry.

\subsection{\label{sec-2nd proced}Second observable}
In the previous section, we tried to find parts of the observable that generate a BLM pattern. It is possible to write the observable in a complete BLM
 pattern form in terms of an appropriate effective charge. This means an absolute freedom in choosing the first two coefficients $e_1$, $e_2$ or
 equivalently $b_1$, $b_2$:
\begin{eqnarray*}\label{CR1.f}
R \; &&= e_1 a_2 \;+\; e_2 a_2^2 \;+\; e_3 a_2^3 \;+\; e_4 a_2^4 \;+\; \ldots \;,\\
a_2 &&= b_1 a \;+\; b_2 a^2 \;+\; b_3 a^3 \;+\; b_4 a^4 \;+\; \ldots \;.
\end{eqnarray*}
As in Eqs.(\ref{r3pre},\ref{r4pre.ss}) we will have the following relations for $e_i$ coefficients:
\begin{eqnarray}\label{CR1}
&&e_3 = -\frac{121}{16}\frac{C_{A}^2}{T_F^2}\frac {e_{21}^2}{e_1}\;-\;\frac{17}{8}\frac{C_{A}^2}{T_F}e_{21} \;-\; \frac{11}{2}\frac{C_{A}}
{T_F}\frac{e_{20} e_{21}}{e_1} \nonumber\\
& & \;\;\;\;+ \;(2 \frac{e_{20} e_{21}}{e_1} \;+\; \frac{5}{4} e_{21} C_{A} \;+\; \frac{3}{4} e_{21} C_{F}) n_f \;+\; \frac{e_{21}^2} {e_{1}} n_f^2 
\nonumber\\
&& e_4 = \frac{C_A}{e_1 T_F} [- \frac {11}{2} e_{20}  e_{31} \;+\; \frac{33}{8} e_{20} e_{21} C_F \;-\; \frac {33}{4} e_{30} e_{21} ] \nonumber\\
& & \;\;\;\; +\; \frac{C_A^3}{T_F^2} [ \frac{627}{32}  \frac{e_{21}^2}{e_1} \;-\;\frac {187}{32}  e_{32} ] \;+\; \frac{C_A^3}{T_F} \frac{203}{1152}  
e_{21} \nonumber\\
& & \;\;\;\; +\; \frac{C_A^2}{T_F} [ \frac{51}{32} e_{21} C_F \;-\; \frac {17}{8} e_{31} ]
+\;  \frac{C_A^2}{e_1 T_F^2} [ \frac{1089}{64} e_{21}^2 C_F \nonumber\\
& & \;\;\;\;-\; \frac{363}{16} e_{21} e_{31} ] \;+\; \ldots ,
\end{eqnarray}
so there is an infinite number of observables in terms of which a BLM pattern is obtained for $R$.

A very first choice for $e_1$, $e_2$ would be $r_1$, $r_2$. For $\tilde{R}$ the predicted result suffers the same problem as the normal BLM prediction
 i.e., a large deviation due to the first term. An impossible state would be the generation of both $e$ and $b$ through BLM pattern.

Obviously we cannot use strict constraints to obtain some $b_1$, $b_2$ because these cannot be partners to the first constraint due to the BLM pattern.
 For instance if $a_2$ is determined as the Adler function corresponding to $R$, $e_1$ will not be able to normalize $R$ appropriately.
 It is enslaved to take $r_1$ while it must play the role of an important building block in the BLM pattern at the same time. Generating all $b$
 coefficients through Adler function except $b_1$ and $b_2$ would be a much more flexible way which reduces the infinite set of $a_2$ observables to a
 set of few $a_2$s.

Every $a_2$ observable connects to $R$ through a specific channel. Temporarily we put the BLM pattern constraint aside. If we consider $a_2$ to be the 
Adler function corresponding to $R$, a very simple dependence of the 4th order coefficient of $R$ to both $e_4$ and $b_4$ is obtained. So either one can
 predict both $e_4$ and $b_4$ or just predict $e_4$ and use Adler transformation for $b_4$ at the same time. Performing the first
 choice for $\tilde{R}$ results in a pattern crossing the exact pattern around $n_f=5$ and for the second choice a crossing at $n_f=2$.

A more reasonable step would be to start with a specific type of relation between two observables and then equip the situation with a BLM pattern. Of
 course it would be a waste of time if this ends to direct relations where we encounter observables whose predictions are equivalent. What we mean by a
 direct relation is a kind of linear relation between two variables. The situation is clarified later. For now we should start with a Crewther like 
relation between two observables $R$ and $a_2$:
\begin{eqnarray}\label{CR2}
[r_0 + r_1 a_s + \sum_{i=2}^\infty r_i {a_s}^i] \; [r_0 - r_1 a_s + \sum_{i=2}^\infty b_i {a_s}^i] \nonumber\\
 = \; C + \frac{\beta_a K}{a},
\end{eqnarray}
where for the first observable we have $R=r_1 a_s + \sum_{i=2}^\infty r_i {a_s}^i$, and for the second one we have $a_2=- r_1 a_s + \sum_{i=2}^\infty b_i
 {a_s}^i$. $\beta_a$ is the famous $\beta$-function $-\sum_{i=0}^\infty \beta_i a^{i+2}$, and $K=\sum_{i=1}^\infty K_i a^{i}$. $\{r_0...r_i\}$ and
 $\{K_1...K_{i-1}\}$ determines any coefficient $b_i$ in a special manner, considering all known coefficients prior to $r_i$ and $K_{i-1}$,
\begin{eqnarray}\label{CR2.direct}
b_{ij} \;\;\;\;\; &&= \; -r_{ij} - \frac{11}{12} \frac{C_A}{r_0} K_{(i-1)j} + C_{ij} \;\;\; j<i-1 \nonumber\\
b_{i(i-1)}  &&= \; -r_{i(i-1)} - \frac{T_F}{3 r_0} K_{(i-1)(i-2)}.
\end{eqnarray}
Despite the fact that $R$ does not contain $r_0$, it plays a crucial role in normalizing $r_i$ coefficients in Eq.(\ref{CR2.direct}). To involve $r_0$ in
 the problem, one should perform prediction one order backward and normalize the result with the appropriate $r_0$. Finally this will make $e_i$
 coefficient be determined in terms of $\{r_1...r_{i-1}\}$. Taking $K$ coefficients as the free variables, Eq.(\ref{CR2.direct}) reduces into the very 
simple direct relation $b_i = - r_i + f_i(n_f)$. The result comes out of such a relation suffers the same problem as a normal prediction for $r_i$ since
 $b_i$ and $r_i$ can exchange roles. The same is true about predicting $r_i$ through $e_i$. Although we are able to determine $e_i$ prior to
 $r_i$ using Eq.(\ref{CR2}), at the same time $e_i$ has a direct relation with $r_i$ i.e., $e_i=c_i r_i + h_i(n_f)$. So if anything is to give us
 real different results, it must have an indirect relation with $r_i$:
\begin{eqnarray*}\label{CR2.indirect}
b_{ij} &&=\sum_{j=0}^{i-1} [l_{ij} r_{ij} + m_{ij} K_{(i-1)(j-1)}] + C_{ij},\\
e_i \; &&=\sum_{j=0}^{i-1} c_{ij} r_{ij} + h_i(n_f).
\end{eqnarray*}
We have done a kind of converting such direct relations into indirect ones in the past when we tried to interpret DPA as changing weights of
 $r_{ij}$ coefficients in the normal BLM pattern. Obviously Eq.(\ref{CR2.direct}) is also a kind of indirect relation between $r_i$ and $K_{i-1}$.
 Following this line, it is clear that involving $i$th order $K$ coefficients in Eq.(\ref{CR2.direct}) instead of $(i-1)$th order ones could be
 obtained by considering $\frac {\beta_a K}{a^2}$ in Eq.(\ref{CR2}):
\begin{widetext}
\begin{eqnarray}\label{CR2.indirect.ext}
&&[r_0 + r_1 a + (r_{20} + r_{21} \beta_0) a^2 + \sum_{i=3}^\infty r_i {a}^i] \; [r_0 - r_1 a + (b_{20} + b_{21} \beta_0) a^2 + \sum_{i=3}^\infty b_i {a}^i] \; = \; C + \frac{\beta_a K}{a^2},
\end{eqnarray}
\end{widetext}
where we've written all the coefficients in the $\beta$-representation so $r_i=r_{(i-1)(i-2)} \beta_{i-2}  +  \sum_{j=0}^{i-1} r_{ij} \beta_0^j$. The
 intrinsic property of this constraint would be to eliminate $\{K_0,K_1,K_{21},K_{32},K_{34}\}$ from the pattern. It is possible to use 
Eq.(\ref{CR2.indirect.ext}) alongside Eq.(\ref{CR2}). When $\frac{\beta_a K}{a}$ get involved in Eq.(\ref{CR2.indirect.ext}), an oversimple indirect
 relation appears. The indirectness comes from the absence of $K_i$ parts in $b_{i0}$ :
\begin{eqnarray}\label{CR2.direct.coeff.ext}
b_{i0} \;\;\;\; &&= \; -r_{i0}  +  D_{ij} \nonumber\\
b_{ij} \;\;\;\; &&= \; -r_{ij} - \frac{K_{i(j-1)}}{r_0}  +  E_{ij} \;\;\; 0<j<i-1 \nonumber\\
b_{i(i-1)}  &&= \; -r_{i(i-1)} - \frac{K_{i(i-2)}}{r_0}.
\end{eqnarray}
As in Eq.(\ref{CR2.direct}) we combined all lower order coefficients' dependences in constant parts $D_{ij}$ and $E_{ij}$. Again prediction will depend
 on $r_0$ which should be resolved like before.

One could rewrite Eq.(\ref{CR2.direct}) in terms of the coefficients of the $\beta$-independent polynomials ${\cal P}_n$ in Eq.(\ref{CR.extension}) using
 the relations in \cite{Kataev:2010du}. The importance of doing this is that the idea in \cite{Kataev:2010du} strongly reveals existence of flavor-
independent polynomials ${\cal P}_n$. Conformal symmetry breaking totally happens through $\beta$-function in this way. So it is more appropriate to
 rewrite Eq.(\ref{CR2.direct}) in terms of these polynomial coefficients, even though this does not affect the directness of Eq.(\ref{CR2.direct}). 

\section{\label{sec-concl}Conclusions}
Flexibility of any optimization prescription in perturbative analysis is the key to check the possibility of improving the prediction by taking  
deviation of lower order predictions into account i.e., finding patterns for the perturbative series. Taking DPA as finding parts of the observable that
 generates the appropriate BLM pattern for each order might help us in categorizing observables and their responses to the prediction respecting their
 deviation patterns.  But it could not be fully utilized to improve the prediction since for any order $n$ we will suffer lack of information about
 $(n-1)$ parameters. In fact the effect of the symmetry group parameters on the BLM pattern is much more noticeable where the overall normalization plays
 a key role. A reasonable way to improve predicted patterns is to perform predictions based on indirect relations. Observables are connected in a
 direct channel when they are expressed in terms of each other. In this case even specific insightful constraints such as the Crewther relation does not
 guarantee a simple indirect connection between observables. As a note for the reader, clearly we are not referring to the physical content of Crewther
 relation but just the mathematical form of it as a connection channel. So finding a mother constraint that produces elegant indirect relations is the
 challenging problem in the context of finding patterns.



\bibliography{FPWQR}

\end{document}